\documentclass[11pt]{revtex4-2}

\usepackage{graphicx}
\usepackage{dcolumn}
\usepackage{bm}

\usepackage{amsmath, amssymb}

\usepackage[utf8]{inputenc}
\usepackage[T1]{fontenc}
\usepackage{mathptmx}
\usepackage{etoolbox}

\newcommand{\Rpm}[1]{\mathcal{R}_{#1}^{(\pm)}}

\begin{document}

\title[Semi-classical evaporation]{Semi-classical evaporative cooling:\\classical and quantum distributions}

\author{A. A. Arvizu-Velazquez}
\affiliation{Centro de Investigaci\'{o}n en Computaci\'{o}n, Instituto Polit\'{e}nico Nacional, Av. Luis Enrique Erro S/N, Unidad Profesional Adolfo L\'{o}pez Mateos, Zacatenco, Alcald\'{i}a Gustavo A. Madero, C.P. 07738, Ciudad de M\'{e}xico.}
\altaffiliation[Also at ]{Facultad de Ciencias, Universidad Nacional Aut\'{o}noma de M\'{e}xico, Alcald\'{i}a Coyoac\'{a}n, C.P. 04510 Ciudad Universitaria, Ciudad de M\'{e}xico, M\'{e}xico}

\author{A. del R\'{i}o-Lima}

\affiliation{
Instituto de Qu\'{i}mica, Universidad Nacional Aut\'{o}noma de M\'{e}xico, Apartado Postal 20-364, 01000 Ciudad de M\'{e}xico, M\'{e}xico
}

\author{S. Dond\'{e}-Rodr\'{i}guez}
\affiliation{
Instituto de F\'{i}sica, Universidad Nacional Aut\'{o}noma de M\'{e}xico, C.P. 04510, M\'{e}xico
}

\author{F. J. Poveda-Cuevas}
\affiliation{
Investigador por M\'{e}xico - Instituto de F\'{i}sica, Universidad Nacional Aut\'{o}noma de M\'{e}xico, C.P. 04510, M\'{e}xico
}

\date{\today}

\begin{abstract}
We develop a semiclassical thermodynamic framework for the evaporative cooling of trapped atomic gases that treats Maxwell--Boltzmann, Bose--Einstein, and Fermi--Dirac statistics on equal footing across box, harmonic, mixed, and linear-quadrupole potentials. Using the global thermodynamic variables of inhomogeneous confinement, we show that all geometries are unified by a single parameter $s$, which fixes the polylogarithm order, the density of states exponent, and the number of degrees of freedom $2s$. Modeling evaporation as a recursive sequence of energy truncation and rethermalization, we derive closed-form recurrence relations for the particle number and internal energy that track the full thermodynamic state, with the classical energy budget set by a virial factor $C_{\mathrm{trap}} = 1 + 3/(2s)$. Quantum degeneracy emerges not as a singularity in the global susceptibilities, but as a smooth, geometry-dependent crossover in which bosons and fermions display opposite thermodynamic signatures. The results provide a versatile theoretical tool for modeling evaporative cooling across experimentally relevant geometries and offer quantitative guidance for optimizing the cooling process in ultracold atomic systems.
\end{abstract}

\maketitle

\global\long\def\Li{\operatorname{Li}}
\global\long\def\erf{\operatorname{erf}}

\section{Introduction}
\label{sec:introduction}

Evaporation in atomic traps is essential for achieving both fermionic \cite{Luo-njp8} and bosonic \cite{Davis-prl74, Anderson-science269, Barrett-prl87} quantum degeneracy. In experimental practice, the main objective is to optimize the balance between evaporation rate and atomic number conservation. The signal-to-noise ratio, which is critical for control experiments and quantum simulators, depends on both the number of particles and the final temperature. Determining optimal evaporation parameters is typically a manual process, as they are highly sensitive to experimental conditions and prior cooling steps, including trap depth, potential geometry, and chamber vacuum. For theoretical analysis, a global thermodynamic description is preferable, since relevant thermodynamic quantities can be measured directly during experimental sequences and used to model the adiabatic process along the evaporation path. In contrast, microscopic approaches \cite{Holland-pra55, Camacho-Guardian-jpb} and kinetic theory \cite{Yamashita-pra59, Luiten-pra53} encounter significant challenges in accurately modeling these processes.

The production of ultracold samples has generated significant interest in the study of quantum phenomena, including atomic collisions, superfluidity, and degenerate quantum gases. Advances in both experimental and theoretical research on ultracold trapped gases have substantially improved the understanding of these thermodynamic systems. This work contributes to ongoing efforts to control and manipulate quantum systems, which have potential technological applications in various fields \cite{Bloch-naturephys8, Bloch-revmodphys80, Hofstetter-jpb51}. The versatility and precise control of external parameters provided by quantum gases are particularly noteworthy \cite{Cornell-revmodphys74, Ketterle-revmodphys74}.

A key aspect of engineering quantum gases is the manipulation of light-matter interactions. Parameters such as light intensity, frequency, phase, and detuning can significantly affect ultracold or quantum gases by enabling the creation of focused optical potentials that serve as conservative traps \cite{Grimm-advatmmol42}. With far-red detuning, atoms experience an attractive force toward the trap center \cite{Barrett-prl87, Stamper-Kurn-prl80, Granade-prl88}, while blue detuning produces a repulsive potential \cite{Davis-prl75, Heo-pra83}. This tunability is fundamental to designing potentials with specific characteristics \cite{Dudley-ProcSPIE4985, Gauthier-optica3}.

Given the capability to experimentally realize a wide range of external potentials \cite{Gaunt-prl110, Mukherjee2017, Busley2022,  Navon2021, FryeArndt2026, Roati2026}, this article examines the following configurations: a three-dimensional (3D) box, a two-dimensional (2D) box combined with a one-dimensional (1D) harmonic oscillator (HO), a 1D box combined with a 2D HO, a 3D HO, and the Quadrupolar potential. The inclusion of mixed configurations is motivated by the significance of external traps that combine harmonic and box-type potentials in advancing two-dimensional physics, as demonstrated for both Bose \cite{Rath-pra82, Yefsah-prl107} and Fermi gases \cite{Martiyanov-prl105, Hueck-prl120}.

A further important consideration is that atomic gases are confined by magnetic and electric fields rather than by rigid-walled vessels. Consequently, the thermodynamic description of the system must be reformulated using new global thermodynamic variables \cite{Romero-Rochin-prl94, Romero-Rochin-bjp35}. These variables define extensive quantities in terms of potential parameters rather than vessel volume, aligning with experimental methodologies \cite{Romero-Rochin-pra85, Poveda-Cuevas-pra92, Mercado-Gutierrez-bjp48}.

In the global thermodynamic framework employed here, the extensive variables associated with trapped gases obey the standard laws of thermodynamics and retain their conventional physical interpretation despite the absence of rigid boundaries. Their formulation provides a consistent basis for describing both the normal and degenerate regimes of confined atomic fluids. A central feature revealed by this approach is that the onset of critical behavior does not manifest through discontinuities in the global susceptibilities or singularities in the derivatives of the free energies--Helmholtz, Gibbs, grand potential, or enthalpy. Instead, the transition is governed by the behavior of the chemical potential and the polylogarithmic structure of the semiclassical distributions, which encode the emergence of degeneracy in inhomogeneous traps.

The article is structured as follows: Section \ref{sec:thermodynamics} provides a brief overview of global thermodynamics and the semiclassical distribution. Section \ref{sec:evaporativecooling} outlines the protocol for evaporating a cloud within a trap. Finally, Section \ref{sec:numericalresults} presents the numerical results and their connections to experimental implementations.

\section{Thermodynamics of trapped atomic gases}
\label{sec:thermodynamics}

\subsection{The Ideal Gas Limit}
\label{subsec:idealgaslimit}

Before the evaporative cooling process, any ultracold atomic gas can be described as an ensemble of $N$ interacting classical particles at sufficiently high temperatures. The complete description is provided by the position and momentum coordinates of each particle $\left\lbrace \mathbf{r}_i,\mathbf{p}_i\right\rbrace$, which defines a point of the system in the six-dimensional phase space.
 
Generally, interactions between particles are characterized by an interatomic central potential $\mathcal{V}\!\left(\mathbf{r}_i-\mathbf{r}_j\right)$. These interactions influence the thermodynamics and dynamics of the gas, thereby determining the time required to reach thermal equilibrium. By comparing characteristic lengths, such as the interaction range~$r_0$ and the interparticle distance~$n_0^{-1/3}$ for a gas with density~$n_0$, one can approximate the effective interaction potential to capture the essential features
of the system.

If the interparticle distance greatly exceeds the interaction range, $n_0\,r_0^3 \ll 1$, the gas is said to be in the dilute regime. In this case, the gas is governed by binary interactions, as the probability of simultaneously finding three atoms within a distance~$r_0$ of each other is extremely low and can be neglected. Furthermore, the collision rate depends on atomic size, as characterized by the cross-section. The kinetic diameter of an atom is approximately equal to the range of the interaction potential, yielding an estimate for the binary cross-section: $\sigma = \pi r_0^2$, which is appropriate for a high-temperature classical regime. The mean free path~$l$ is defined via $l^{-1} = n\sigma$, and the collision rate is
\begin{equation}
  \tau_{c}^{-1} = n\sigma\,\bar{v}_r ,
\end{equation}
where $\bar{v}_r = \sqrt{16k_{B}T/\pi m}$ denotes the average relative atomic
speed, for a Maxwell-Boltzmann distribution of identical particles with mass $m$.
 
By comparing the atomic mean free path with the size of the gas cloud, two density regimes are distinguished: a low-density regime in which the mean free path exceeds the size of the cloud ($l \gg V^{1/3}$), and a high-density regime in which the cloud size exceeds the mean free path ($l \ll V^{1/3}$). In the low-density regime, the gas behaves as a collisionless system, while in the high-density regime, it exhibits hydrodynamic properties.
 
Experimental control over atomic conditions allows the system to enter the collisionless regime.
At the onset of evaporative cooling, corresponding to higher temperatures, the kinetic energy of the collisionless gas dominates the total energy, so that particle interactions may be neglected as a first approximation.
As the temperature decreases and the system transitions to a low-energy regime, interactions become significant and are characterized by the scattering length~$a_s$, which is tunable via an external magnetic field.
By choosing sufficiently weak interaction strengths, the system can be placed in a weakly interacting (nearly ideal) gas regime. Under these conditions, the evaporative cooling process can be analyzed within an
approximation that neglects interactions. The total energy of the gas is then described by the Hamiltonian \cite{Courteille-laserphys11}:
\begin{equation}
  H = \sum_{i}\left(\frac{p_i^2}{2m} + U\!\left(\mathbf{r}_{i}\right)\right),
  \label{eq:002}
\end{equation}
where $p_{i}^{2}/2m$ denotes the kinetic energy of atom~$i$ with mass~$m$,
$p_{i} = |\mathbf{p}_{i}|$ is the magnitude of the particle's momentum, and
$U\!\left(\mathbf{r}_{i}\right)$ is the potential energy associated with the
external trapping field.
 
Equation~(\ref{eq:002}) provides an expression for systems containing many
particles in a macrostate.
In the semiclassical limit, the Hamiltonian is approximated by its classical form:
\begin{equation}
  \epsilon = H\!\left(\mathbf{r},\mathbf{p}\right)
           = \frac{p^2}{2m} + U\!\left(\mathbf{r}\right),
  \label{eq:003}
\end{equation}
where the kinetic and potential energies are averaged over the $N$ particles.
This approach enables the derivation of semiclassical phase-space distribution
functions for classical and quantum gases:
\begin{equation}
  f^{\left(0\right)}\!\left(\epsilon;\mu,T\right)
  = \exp\!\left\{-\frac{1}{k_B T}\!\left[\frac{p^{2}}{2m}
    + U\!\left(\mathbf{r}\right) - \mu\right]\right\},
  \label{eq:004}
\end{equation}
\begin{equation}
  f^{\left(-\right)}\!\left(\epsilon;\mu,T\right)
  = \frac{1}{\exp\!\left\{\dfrac{1}{k_B T}\!\left[\dfrac{p^{2}}{2m}
    + U\!\left(\mathbf{r}\right) - \mu\right]\right\} - 1},
  \label{eq:005}
\end{equation}
\begin{equation}
  f^{\left(+\right)}\!\left(\epsilon;\mu,T\right)
  = \frac{1}{\exp\!\left\{\dfrac{1}{k_B T}\!\left[\dfrac{p^{2}}{2m}
    + U\!\left(\mathbf{r}\right) - \mu\right]\right\} + 1},
  \label{eq:006}
\end{equation}
where $k_B$ denotes the Boltzmann constant.
Equation~(\ref{eq:004}) represents the Maxwell-Boltzmann~(MB) distribution, which
applies to classical systems.
Equations~(\ref{eq:005}) and~(\ref{eq:006}) correspond to the Bose-Einstein~(BE)
and Fermi-Dirac~(FD) distributions, respectively, which describe quantum systems.
All three distribution functions are determined by the chemical potential~$\mu$ and
the system temperature~$T$.
 
\subsection{Semiclassical distribution for trapped gases}
\label{subsec:semiclassical_trapped}

The thermodynamic properties are described within the theoretical framework
presented in~\cite{Sandoval-Figueroa-pre78}, where a generalized volume
$\mathcal{V}$ corresponds to the spatial extension of the gas.
Experimentally, the geometry of the trapped cloud is determined solely by the
characteristics of the trapping potential.
Employing a semiclassical distribution aims to reproduce the functional forms
reported in the literature while maintaining experimental relevance.
Consequently, the calculation of thermodynamic quantities can be reduced to
determining two key variables: the number of particles and the internal energy.
The distribution function $f^{\left(a\right)}\!\left(\mathbf{r},\mathbf{p}\right)$
is advantageous because it can be integrated directly to obtain thermodynamic
variables of either classical or quantum origin~\cite{Giorgini-jlowtphys109}.
 
Using the semiclassical density distributions from Eqs.~(\ref{eq:004})
to~(\ref{eq:006}), it is possible to obtain thermodynamic quantities such as the
number of particles $N$ and the internal energy $E$ for specified values of $\mu$
and $T$. The analysis begins by considering a phase-space cell
\begin{equation}
  dN = \frac{1}{h^3}\,d^3r\;d^3p\;
       f^{\left(a\right)}\!\left(\mathbf{r},\mathbf{p}\right),
  \label{eq:007}
\end{equation}
and a distribution function $f^{\left(a\right)}\!\left(\epsilon;\mu,T\right)$,
which can take the form of Eq.~(\ref{eq:004}) for a classical gas ($a=0$),
Eq.~(\ref{eq:005}) for bosons ($a=-$), or Eq.~(\ref{eq:006}) for
fermions ($a=+$). To simplify the notation, we write
\[
  f^{\left(a\right)}\!\left(\mathbf{r},\mathbf{p};\mu,T\right)
  \equiv f^{\left(a\right)}\!\left(\epsilon;\mu,T\right)
  \equiv f^{\left(a\right)}\!\left(\mathbf{r},\mathbf{p}\right).
\]
This function represents the phase-space occupation at the point
$\left(\mathbf{r},\mathbf{p}\right)$ within an elementary phase-space volume
$\left(2\pi\hbar\right)^{-3}$.
Integrating over the full phase space yields the total number of particles
described by the distribution:
\begin{equation}
  N = \frac{1}{h^{3}}
      \int d^3r \int d^3p\;
      f^{\left(a\right)}\!\left(\mathbf{r},\mathbf{p}\right).
  \label{eq:008}
\end{equation}
Integrating the phase-space density over momentum space gives the probability of
finding an atom at position $\mathbf{r}$, that is, the spatial density
distribution of the gas:
\begin{equation}
  n\!\left(\mathbf{r}\right)
  = \frac{1}{h^{3}}
    \int d^3p\;
    f^{\left(a\right)}\!\left(\mathbf{r},\mathbf{p}\right).
  \label{eq:009}
\end{equation}
Trapped clouds exhibit strong inhomogeneity, as the density decreases from its
maximum at the center of the cloud to the minimum imposed by the external
potential. Consequently, a system confined in an external trap possesses a non-homogeneous density distribution with a profile uniquely determined by the geometry of the confinement.
This behavior becomes observable only when quantum statistics are significant,
which depends strongly on the temperature and the number of atoms.
Experimentally, particularly in ultracold atom physics, direct access to the
three-dimensional sample density $n\!\left(\mathbf{r}\right)$ is not available.
Instead, measurements typically yield the density integrated along one spatial
direction, as a result of standard analysis methods such as saturated absorption
imaging~\cite{Ramanathan-revsciinst83,Szczepkowski-revsciins80}.
Nevertheless, it is possible to distinguish between two-dimensional density
profiles, each exhibiting a unique signature.
Using Eq.~(\ref{eq:009}), the number of particles can be expressed as a function
of the spatial density distribution:
\begin{equation}
  N\!\left(\mu,T\right) = \int d^3r\; n\!\left(\mathbf{r}\right).
  \label{eq:010}
\end{equation}
Thermodynamically, the canonical relationship between conjugate variables is
established through the product of the chemical potential and the number of
particles.
This relationship implies that $N\!\left(\mu, T\right)$ is an extensive quantity, a property that must be preserved by definition. Therefore, even though the sample is distributed inhomogeneously due to the external potential, the functional form of the particle number must be consistent with this extensivity.
 
Similarly, for both classical and quantum distributions in the ideal-gas limit, the internal energy of the system is obtained by integrating the single-particle energy over phase space:
\begin{equation}
  E\!\left(\mu,T\right)
  = \frac{1}{h^{3}}\int d^{3}r\int d^{3}p
    \left(\frac{p^{2}}{2m}+U\!\left(\mathbf{r}\right)\right)
    f^{\left(a\right)}\!\left(\mathbf{r},\mathbf{p}\right).
  \label{eq:011}
\end{equation}
Likewise, energy is an extensive quantity by definition, and its explicit form depends on the external potential. This expression provides a method for calculating the mean total energy.

At this stage, the classical gas can be distinguished from the quantum gas by specifying the respective distribution functions, integrating over momentum space, and retaining the explicit functional dependence on chemical potential, temperature, and trapping potential.
For the classical gas, this yields
\begin{equation}
  N\!\left(\mu,T\right)
  = \int d^{3}r\,\frac{\mathrm{e}^{\alpha\!\left(\mathbf{r}\right)}}{\lambda^{3}}
  \label{eq:012}
\end{equation}
and
\begin{equation}
  E\!\left(\mu,T\right)
  = \frac{3\pi\hbar^{2}}{m}
    \int d^{3}r\,
    \frac{\mathrm{e}^{\alpha\!\left(\mathbf{r}\right)}}{\lambda^{5}}
    + \int d^{3}r\,U\!\left(\mathbf{r}\right)
      \frac{\mathrm{e}^{\alpha\!\left(\mathbf{r}\right)}}{\lambda^{3}},
  \label{eq:013}
\end{equation}
where $\lambda = \left(h^{2}/2\pi mk_{B}T\right)^{1/2}$ is the thermal de~Broglie wavelength, and the local fugacity is given by 
\begin{equation}
  \mathrm{e}^{\alpha\!\left(\mathbf{r}\right)}
  = \exp\!\left[\frac{\mu - U\!\left(\mathbf{r}\right)}{k_{B}T}\right].
  \label{eq:014}
\end{equation}
This formulation is based on the local density approximation (LDA), which assumes that spatial density variations are slow compared to the local de~Broglie Wavelength, typically due to smooth external confinement. Under this approximation, the local density is treated as that of a locally uniform system characterized by an effective chemical potential $\mu_{\mathrm{eff}}\!\left(\mathbf{r}\right) = \mu - U\!\left(\mathbf{r}\right)$.

An analogous procedure applied to quantum gases yields the number of particles
\begin{equation}
  N\!\left(\mu,T\right)
  = \int d^{3}r\,
    \frac{g_{3/2}^{\left(\pm\right)}\!\left(\alpha\!\left(\mathbf{r}\right)\right)}
         {\lambda^{3}}
  \label{eq:015}
\end{equation}
and the internal energy
\begin{equation}
  E\!\left(\mu,T\right)
  = \frac{3\pi\hbar^{2}}{m}
    \int d^{3}r\,
    \frac{g_{5/2}^{\left(\pm\right)}\!\left(\alpha\!\left(\mathbf{r}\right)\right)}
         {\lambda^{5}}
    + \int d^{3}r\,U\!\left(\mathbf{r}\right)
      \frac{g_{3/2}^{\left(\pm\right)}\!\left(\alpha\!\left(\mathbf{r}\right)\right)}
           {\lambda^{3}},
  \label{eq:016}
\end{equation}
where $g_{s}^{\left(\pm\right)}$ is the polylogarithm of order~$s$, with superscript $(+)$ for a Fermi-Dirac distribution, and superscript $(-)$ for a Bose-Einstein distribution. For both cases, it is defined by the integral representation
\begin{equation}
  g_{s}^{\left(\pm\right)}\!\left(\alpha\right)
  = \frac{1}{\Gamma\!\left(s\right)}
    \int_{0}^{\infty}dx\,
    \frac{x^{s-1}}{\mathrm{e}^{x-\alpha}\mp 1}.
  \label{eq:017}
\end{equation}
Within this formalism, the primary challenge is to determine the appropriate form
of $U\!\left(\mathbf{r}\right)$. As demonstrated in Refs.~\cite{Bagnato-pra35,Bagnato-pra44}, the potential need not be restricted to polynomial forms; combinations that yield mixed degrees of freedom are also admissible and give rise to distinct thermodynamic behavior.

\subsection{Thermodynamic variables for confining potentials}
\label{subsec:thermoconfining}

 A consequence of the inhomogeneous confinement in quantum gas experiments is that the conventional thermodynamic variables of pressure and volume are no longer well defined, since no rigid walls or containers bound the system. In this case, the pressure becomes intrinsically local, varying with position rather than admitting a single global value, and the volume cannot be assigned unambiguously because the gas has no sharply defined spatial extent. However, a set of global thermodynamic variables that correctly describe the mechanical equilibrium between the external forces and the local pressure gradients can be defined, and this set depends on the characteristics of each confining potential \cite{Romero-Rochin-bjp35, Romero-Rochin-prl94, Sandoval-Figueroa-pre78}. 
 
The confining external potentials analyzed in this work are summarized in Table~\ref{tab:traps}. Each confining potential is associated with a generalized volume variable $\mathcal{V}$ (see Table~\ref{tab:traps}), which may not correspond to a conventional volume unit but is treated as an extensive quantity, as it characterizes the spatial extent of the gas within the atomic trap. The boundaries of the box potentials in different dimensions are given by
\begin{equation}
U_{\mathrm{1D}}=\left\{ \begin{array}{c}
0\quad r\in L\\
\infty\quad\notin L
\end{array}\right. ,
\quad
U_{\mathrm{2D}}=\left\{ \begin{array}{c}
0\quad r\in \Sigma \\
\infty\quad\notin \Sigma
\end{array}\right. ,
\quad
U_{\mathrm{3D}}=
\left\{ \begin{array}{c}
0\quad r\in V\\
\infty\quad\notin V
\end{array}\right. .
\label{eq:018}
\end{equation}
Accordingly, each box potential is defined over a line $L$, a surface $\Sigma$, or a
volume $V$, respectively. Similarly, harmonic confinement can be realized in one, two, or three dimensions.
A primary focus of this work is the anisotropic three-dimensional harmonic
potential, characterized by $\omega_x \neq \omega_y \neq \omega_z$, as most
experimentally implemented potentials belong to this category.
Potential designs relevant to homogeneous gases have also motivated this study.
Various experimental realizations have employed mixed potentials that combine
harmonic and box-type features.
These configurations have been instrumental in advancing two-dimensional physics,
as demonstrated for both Bose~\cite{Rath-pra82,Yefsah-prl107} and Fermi
gases~\cite{Martiyanov-prl105,Hueck-prl120}.
The combination of harmonic and box potentials is also considered here, with the
respective generalized volumes listed in Table~\ref{tab:traps}.

\begin{table}[htbp]
\centering
\caption{Definition and parameters characterizing the confining external potentials $U(\mathbf{r})$. The parameter $\mathcal{V}$ denotes the generalized volume associated with each potential. The parameter $K$, referred to as the \textit{mass factor}, arises in the definition of the number of particles given in~(\ref{eq:019}). The factor $s$ denotes the order of the poly-logarithm function $g_s^{(\pm)}$ defined in~(\ref{eq:017}), and is related to the total number of degrees of freedom imposed by the external potential through $2s$. Because $s$ simultaneously fixes the poly-logarithm order, the density-of-states exponent, and the power of $T$ in the particle-number prefactor of~(\ref{eq:019}), it acts as the single parameter that unifies the thermodynamic description across all geometries considered.}
\label{tab:traps}
\begin{tabular}{lcccc}
\hline
Potential & $U(\mathbf{r})$ & $\mathcal{V}$ & $K$ & $s$ \\
\hline
3D-Box & $U_{\mathrm{3D}}\left(x,y,z\right)$ & $V$ & $\left(\frac{m}{2\pi}\right)^{3/2}$ & $3/2$ \\[6pt]
2D-Box + 1D-HO & $U_{\mathrm{2D}}\left(x,y\right)+\frac{1}{2}m\omega_{z}^2z^2$ & $\Sigma \left(\omega_z\right)^{-1}$ & $\left(\frac{m}{2\pi}\right)^{2}$ & $2$ \\[6pt]
1D-Box + 2D-HO & $U_{\mathrm{1D}}\left(z\right)+\frac{1}{2}m\left(\omega_{x}^2x^2+\omega_{y}^2y^2\right)$ & $L\left(\omega_x\omega_y\right)^{-1}$ & $\left(\frac{m}{2\pi}\right)^{5/2}$ & $5/2$ \\[6pt]
3D-HO & $\frac{1}{2}m\left(\omega_{x}^2x^2+\omega_{y}^2y^2+\omega_{z}^2z^2\right)$ & $\left(\omega_{x}\omega_{y}\omega_{z}\right)^{-1}$ & $1$ & $3$ \\[6pt]
Quadrupole & $\sqrt{\left(A_x x\right)^{2}+\left(A_y y\right)^{2}+\left(A_z z\right)^{2}}$ & $\left(A_x A_y A_z\right)^{-1}$ & $8\pi\left(\frac{m}{2\pi}\right)^{3/2}$ & $9/2$ \\
\hline
\end{tabular}
\end{table} 

The concept of the number of degrees of freedom, given by $(2s)$ (see
Table~\ref{tab:traps}), with $s$ as the order of the
poly-logarithm function $g_s^{(\pm)}$ defined in~\eqref{eq:017}, warrants further discussion.
For the purely box-confined gas, there are three spatial degrees of freedom.
As additional constraints are introduced through the confinement field,
specifically the curvatures associated with the harmonic potential, $s$
increases, reaching its maximum in the purely harmonic case. Harmonic confinement thus effectively doubles the number of degrees of freedom compared to box confinement (a free gas in a box with hard walls). A notable special case is the linear quadrupole trap, which is widely employed in experiments that initiate evaporation in a purely magnetic trap.
Experimentally, this configuration presents technical challenges owing to the
presence of a zero-field point and the associated loss of colder atoms.
Nevertheless, at high temperatures, this potential offers a larger phase-space
volume, facilitating more rapid thermalization during evaporative cooling.
The geometric scale of the trap is typically set by the gradient of the magnetic
field ($B^{\prime}$) along its strongest direction:
\[
  \mu_B B^{\prime} = A_x = A_y = 2A_z,
\]
where $\mu_B$ denotes the Bohr magneton.
In this potential, the physical interpretation of the degrees of freedom becomes
ambiguous; nevertheless, $(2s)$ will continue to be used as a label for the effective
number of degrees of freedom, despite the loss of direct physical correspondence.
 
Both the number of particles and the energy of classical and quantum gases can be
determined analytically. These expressions serve as the primary tool for illustrating the differences that arise as the system approaches the degeneracy temperature.
The general expressions for the number of particles are
\begin{equation}
  N = K\,\frac{\mathcal{V}}{\hbar^3}\left(k_{B}T\right)^{s}\mathrm{e}^{\alpha},
  \qquad
  N = K\,\frac{\mathcal{V}}{\hbar^3}\left(k_{B}T\right)^{s}
      g^{\left(\pm\right)}_{s}\!\left(\alpha\right),
  \label{eq:019}
\end{equation}
for the classical and quantum gases, respectively, where $K$, the \textit{mass factor}, is listed in
Table~\ref{tab:traps} for each potential considered.
The corresponding expressions for the internal energy are
\begin{equation}
  E = s N k_B T,
  \qquad
  E = s N k_B T\,
      \frac{g^{\left(\pm\right)}_{s+1}\!\left(\alpha\right)}
           {g^{\left(\pm\right)}_{s}\!\left(\alpha\right)}.
  \label{eq:020}
\end{equation}
The equipartition theorem states that the mean internal energy per particle is $E/N = (2s) k_B T/2$, where $2s$ denotes the degrees of freedom. This expression holds only for potentials that are quadratic (harmonic) or have infinite walls (box). For the quadrupole potential, $U(\mathbf{r}) \propto |\mathbf{r}|$, the virial theorem yields a different kinetic-to-potential energy ratio.

As stated in this subsection, the global thermodynamic variables of volume and pressure can be defined, and through them, the number of particles and the internal energy. Thus, as in the traditional framework of thermodynamics, it is possible to compute all the thermodynamic properties characterizing the system, such as the Helmholtz and Gibbs free energies, the grand potential, the enthalpy, the susceptibilities, etc. Of particular importance are the susceptibilities, which allow for the description of phase transitions. 

\subsection{Free energies and susceptibilities for confining potentials}
\label{subsec:freeenergies}

As stated in the previous subsection, global thermodynamic variables analogous to volume and pressure can be consistently defined for an inhomogeneously confined gas. Together with the particle number and internal energy given in equations~\eqref{eq:019} and~\eqref{eq:020}, these variables enable a complete thermodynamic description of the system. This description is naturally formulated in the grand canonical ensemble, in which the temperature $T$, generalized volume $\mathcal{V}$, and chemical potential $\mu$ are the independent variables. The corresponding thermodynamic potential is
\begin{equation}
\Omega\left(\mu, T,\mathcal{V}\right) = E - TS - \mu N = -\mathcal{P}\mathcal{V},
\label{eq:grandpotential}
\end{equation}
where $\mathcal{P}$ is the generalized pressure conjugate to $\mathcal{V}$. It shows that the entropy, generalized pressure, and particle number can be obtained directly from the first derivatives of $\Omega$. Once the grand potential is determined, the other thermodynamic potentials follow from standard Legendre transformations, thereby establishing the full equilibrium thermodynamics of the confined gas \cite{Fermi1956, Scherer1990}.

The resulting closed-form expressions for the entropy $S\left(N, E, \mathcal{V}\right)$, enthalpy $H\left(S, \mathcal{P}, N\right)$, Helmholtz free energy $F\left(N,T,\mathcal{V}\right)$, and Gibbs free energy $G\left(\mathcal{P}, T, N\right)$ are summarized in Table~\ref{tab:potentials} for both classical and quantum statistics. In the classical regime, these quantities recover the Maxwell--Boltzmann results, whereas for Bose--Einstein and Fermi--Dirac gases they are expressed in terms of ratios of the functions in compact form as:
\begin{equation}
  \mathcal{R}_{s}^{(\pm)}(\alpha)
  \equiv
  \frac{g_{s}^{(\pm)}(\alpha)}{g_{s-1}^{(\pm)}(\alpha)}.
  \label{eq:polylog-ratio}
\end{equation}
The dependence on the confining geometry is incorporated through the parameter $s$ listed in Table~\ref{tab:traps}. Consequently, the same formalism applies uniformly to all trapping potentials considered here.

\begin{table}[htbp]
\centering
\caption{Closed-form expressions for the principal thermodynamic potentials of a trapped ideal gas in the ideal-gas limit. The upper (lower) sign denotes Bose--Einstein (Fermi--Dirac) statistics, and $\alpha=\mu/(k_BT)$. The classical expressions are recovered in the non-degenerate limit.}
\label{tab:potentials}
\begin{tabular}{lll}
\hline
 & Classical & Quantum \\
\hline
Entropy ($S$) & $\displaystyle Nk_B\left[\ln\!\left(\dfrac{\mathcal{V}}{N\lambda^{3}}\right)+s+1\right]$ & $\displaystyle Nk_B\left[(s+1)\mathcal{R}_{s+1}^{(\pm)}-\alpha\right]$ \\[12pt]
Enthalpy ($H$) & $\displaystyle (s+1)Nk_BT$ & $\displaystyle (s+1)Nk_BT\mathcal{R}_{s+1}^{(\pm)}$ \\[12pt]
Helmholtz ($F$) & $\displaystyle -Nk_BT\left[\ln\!\left(\dfrac{\mathcal{V}}{N\lambda^{3}}\right)+1\right]$ & $\displaystyle Nk_BT\left[\alpha-\mathcal{R}_{s+1}^{(\pm)}\right]$ \\[12pt]
Gibbs ($G$) & $\displaystyle -Nk_BT\ln\!\left(\dfrac{\mathcal{V}}{N\lambda^{3}}\right)$ & $\displaystyle Nk_BT\alpha$ \\
\hline
\end{tabular}
\end{table}

Beyond the thermodynamic potentials themselves, a complete characterization of the system requires the response functions associated with variations of temperature, generalized volume, and generalized pressure. Table~\ref{tab:responses} presents the specific heat capacities at constant generalized volume and pressure, 
\begin{equation}
    c_{\mathcal{V}} = - T\left(\frac{\partial^2 f}{\partial T^2}\right)_{\mathcal{V}/N} , \quad c_{\mathcal{P}} = \left(\frac{\partial h}{\partial T}\right)_{\mathcal{P}},
\end{equation}
together with the global isothermal compressibility and global thermal expansion coefficient
\begin{equation}
    \mathcal{K}_{T} = -\left(\frac{\partial \mathcal{P}}{\partial g}\right)_T \left(\frac{\partial^2 g}{\partial \mathcal{P}^2}\right)_T, \quad \mathcal{B}_{\mathcal{P}} = \left(\frac{\partial \mathcal{P}}{\partial g}\right)_T \left(\frac{\partial^2 g}{\partial T \partial\mathcal{P}}\right)_T,
\end{equation}
respectively. $h$, $f$, and $g$ are the thermodynamic potentials in its specific form. These susceptibilities are derived from thermodynamic potentials and quantify the response of the trapped gas to infinitesimal changes in its thermodynamic state. In the classical limit, they reduce to the corresponding Maxwell--Boltzmann expressions, while in the quantum regime their polylogarithmic dependence captures the departure from classical behavior as degeneracy is approached.

The susceptibilities are particularly relevant because they provide direct thermodynamic signatures of the normal-to-degenerate transition. Their behavior distinguishes the response of Bose and Fermi gases and reveals how the trapping geometry modifies the onset and magnitude of quantum-statistical effects. Thus, the thermodynamic variables, together with the thermodynamic potentials, provide the complete set of quantities required to follow the thermodynamic state of the gas throughout the evaporative cooling protocol. 


\begin{table}[htbp]
\centering
\caption{Linear response functions for classical and quantum statistics. The shorthand $\mathcal{R}_{j}^{(\pm)}\equiv\mathcal{R}_{j}^{(\pm)}(\alpha)$ is defined in equation~\eqref{eq:polylog-ratio}.}
\label{tab:responses}
\begin{tabular}{lll}
\hline
 & Classical & Quantum \\
\hline
Specific heat capacity ($c_{\mathcal{V}}$) & $\displaystyle s$ & $\displaystyle s(s+1)\Rpm{s+1}-s^{2}\Rpm{s}$ \\[12pt]
Specific heat capacity ($c_{\mathcal{P}}$) & $\displaystyle s+1$ & $\displaystyle (s+1)\frac{\Rpm{s+1}}{\Rpm{s}}\times\left[(s+1)\Rpm{s+1}-s\Rpm{s}\right]$ \\[12pt]
Isothermal compressibility ($\mathcal{K}_{T}$) & $\displaystyle \frac{1}{\mathcal{P}}$ & $\displaystyle \frac{\mathcal{V}}{Nk_BT\,\Rpm{s}}$ \\[12pt]
Thermal expansion ($\mathcal{B}_{\mathcal{P}}$) & $\displaystyle \frac{1}{T}$ & $\displaystyle \frac{(s+1)\Rpm{s+1}-s\Rpm{s}}{T\,\Rpm{s}}$ \\
\hline
\end{tabular}
\end{table}

\section{Evaporative cooling}
\label{sec:evaporativecooling}

In the study of ultracold matter, evaporative cooling~\cite{Davis-prl75, Hess-1986}
is among the most widely used experimental techniques for achieving temperatures
below the recoil and sub-Doppler limits in dilute gases. This technique reduces the depth of the external potential, thereby selectively removing atoms in high-energy states of the thermal distribution. Collisions among the remaining atoms facilitate rethermalization of the system. Given its significance, several theoretical studies have examined this procedure~\cite{Henn-amjphys75, Davis-appphysb60}.

Evaporative cooling is modeled as a discrete process arising from truncation of the energy distribution, which is typically assumed to follow the Maxwell-Boltzmann (MB) distribution. The present study considers not only the MB distribution but also the quantum Fermi-Dirac (FD) and Bose-Einstein (BE) distributions, enabling analysis of variations in thermodynamic behavior during evaporation, including at equilibrium. As detailed below, the three distributions differ significantly at low temperatures. Conversely, at high temperatures, all three gases recover behavior consistent with that of a classical system. For bosonic systems, the critical temperature for achieving quantum degeneracy depends on the external confining potential. Particular attention is given to the quadrupole potential, which exhibits distinct behavior across the three distributions examined. The main results of the theoretical model are presented in the following sections.

\subsection{Evaporative cooling protocol}
\label{subsec:evapcoolingprotocol}

\begin{figure}[htpb!]
    \centering
    \includegraphics[width=0.5\linewidth]{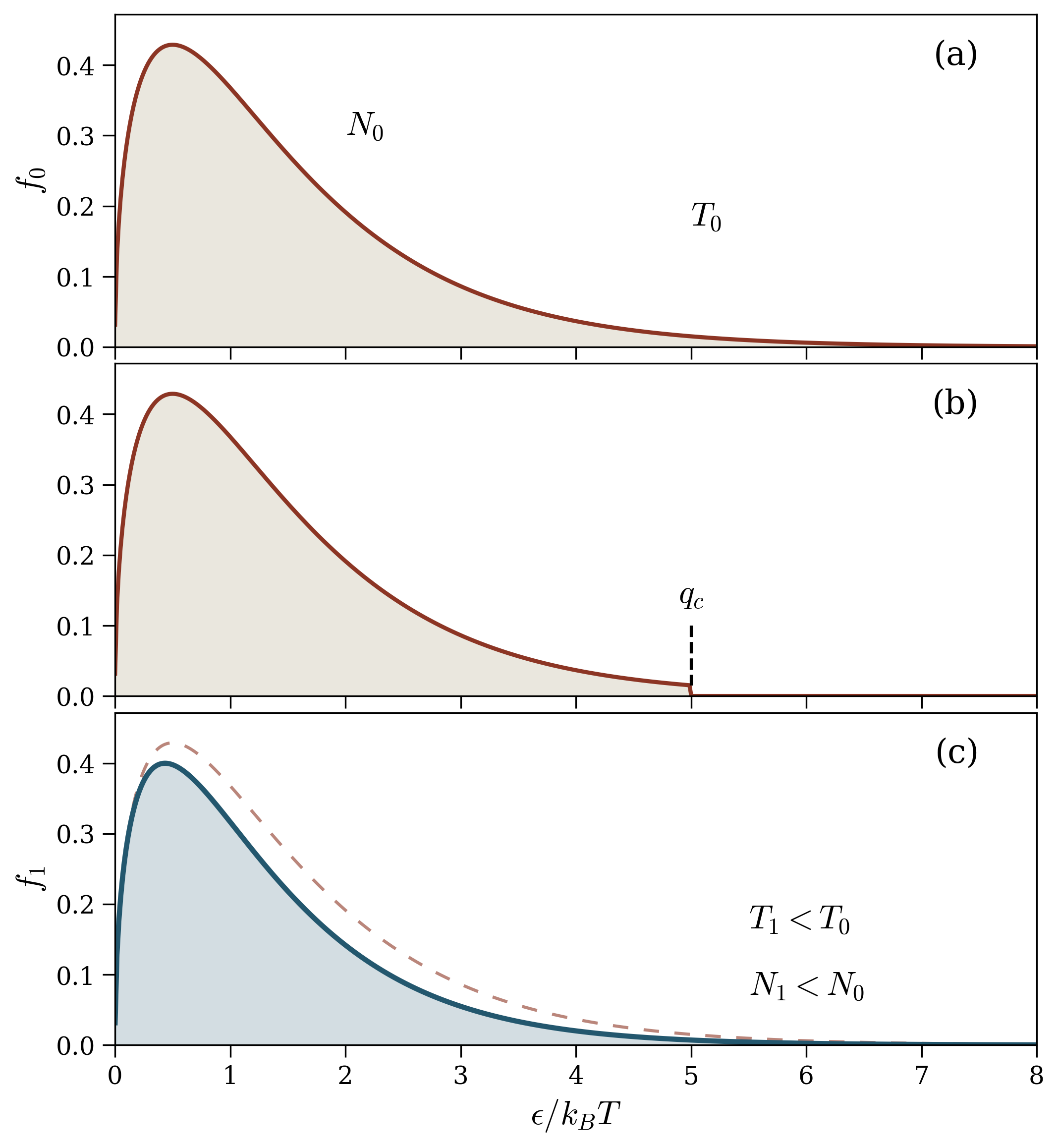}
    \caption{Schematic of a single evaporation step. (a) The cloud starts in equilibrium at temperature $T_0$ with $N_0$ atoms, described by the energy distribution $f_0$. (b) A truncation is imposed at the cut-off momentum $q_c$ (energy $\epsilon_c=q_c^2/2m$), removing the high-energy tail of the distribution. (c) After collisional rethermalization, the gas relaxes into a new equilibrium distribution $f_1$ at a lower temperature $T_1<T_0$ and reduced atom number $N_1<N_0$; the dashed curve reproduces the original $f_0$ for comparison. Iterating this cut-and-rethermalize cycle defines the recursive cooling protocol used throughout this work.}
    \label{fig:fevap}
\end{figure}

The initial particle number, energy, and sample temperature can be determined directly from the distributions presented in equations (\ref{eq:004})--(\ref{eq:006}). The functional form of these distributions depends on the external potential $U\left(\mathbf{r}\right)$, which is embedded in $\epsilon$, defined in~(\ref{eq:003}). The initial temperature is denoted by the subscript $0$. Given a temperature $T_{0}$ and a chemical potential $\mu_{0}$, it is possible to calculate both the number of particles and the energy as follows:
\begin{align}
   N_{0} & =\frac{1}{h^{3}}\int d^{3}r\int d^{3}p\,f_{0}^{\left(a\right)}\left(\mu_0,T_0\right),\label{eq:021}\\
   E_{0} & =\frac{1}{h^{3}}\int d^{3}r\int d^{3}p\,\left(\frac{p^{2}}{2m}+U\left({\bf r}\right)\right)f_{0}^{\left(a\right)}\left(\mu_0,T_0\right).\label{eq:022}
\end{align}
where $a = 0, \pm$ for the MB, BE, and FD distributions, respectively. Note that these expressions are general. Given a distribution, a cut can be performed, establishing a cut-off momentum $q_c$. All particles with momentum greater than $q_c$ are removed. The cut-off momentum corresponds to an energy $\epsilon_c = q_{c}^2/2 m$. The cut-off is chosen so as to maintain adiabaticity and avoid abrupt changes, since the analysis concerns systems near equilibrium at each stage. Otherwise, the discussion would shift to non-equilibrium phenomena, which are beyond the scope of this article. Figure \ref{fig:fevap} schematically represents this process. Initially, before particle removal, the gas is in equilibrium, characterized by the variables $\left(N_0, E_0 \right)$ (Fig. \ref{fig:fevap}(a)). After the cut-off (Fig. \ref{fig:fevap}(b)), the system is described by the new variables $\left(N_1^c, E_1^c\right)$ (Fig. \ref{fig:fevap}(c)), which depend on the initial conditions $T_0$ and $\mu_0$. The distribution is thus redefined by truncating the original momentum-space integral, effectively removing the highest-energy atoms.

The procedure for truncating the integrals is analogous for both the classical and quantum distributions. The truncation of the particle-number integral is first demonstrated for the Maxwell-Boltzmann (MB) distribution:
\begin{eqnarray}\label{eq:023}
N_{1}^{c}\left(\mu_{0},T_{0}\right)  & = & 
      \frac{4\pi}{h^{3}}\int d^{3}r\int_{0}^{q_{c}}dp\,p^{2}f_{0}^{\left(0\right)}\left(\epsilon;\mu_{0},T_{0} \right) \nonumber \\
 &=&\left\{ \mathrm{erf}\left[\sqrt{\eta_{0}}\right]-\frac{2}{\sqrt{\pi}}\sqrt{\eta_{0}}\mathrm{e}^{-\eta_{0}}\right\} \int d^{3}r\,\frac{\mathrm{e}^{\alpha_{0}\left({\bf r}\right)}}{\lambda_{0}^{3}},
\end{eqnarray}
while the corresponding expression for the quantum distributions is
\begin{eqnarray}\label{eq:024}
 N_{1}^{c}\left(\mu_{0},T_{0}\right)   & = &  
    \frac{4\pi}{h^{3}}\int d^{3}r\int_{0}^{q_{c}}dp\,p^{2} f_{0}^{\left(\pm\right)}\left(\epsilon;\mu_{0},T_{0}\right) \nonumber \\
    &=&\frac{1}{\lambda_{0}^{3}}\sum_{j=1}^{\infty}\left(\pm 1\right)^{j+1}\int d^{3}r\,\frac{{\rm e}^{j\alpha_{0}\left({\bf r}\right)}}{j^{3/2}}\mathrm{erf}\left[\sqrt{j\eta_{0}}\right] \nonumber\\
    &-& \frac{q_{c}}{\pi\hbar}\frac{1}{\lambda_{0}^{2}}\sum_{j=1}^{\infty}\left(\pm 1\right)^{j+1}\int d^{3}r\,\frac{{\rm e}^{j\left(\alpha_{0}\left({\bf r}\right)-\eta_{0}\right)}}{j},
\end{eqnarray}
where
\begin{equation}\label{eq:025}
\alpha_{0}\left(\mathbf{r}\right)=\frac{1}{k_{B} T_{0}} \left(\mu_{0}-U\left(\mathbf{r}\right)\right),\quad \eta_{0}=\frac{q_{c}^{2}}{2 m k_{B} T_{0}},
\end{equation}
and
\begin{equation}
    \mathrm{erf}\left[\sqrt{\eta}\right]=\frac{2}{\sqrt{\pi}}\int_{0}^{\sqrt{\eta}}dt\,\mathrm{e}^{-t^{2}},
\end{equation}
where the final expression defines the error function. On the other hand, the energies associated with the truncated distributions are, for the classical case,
\begin{eqnarray}\label{eq:026}
  & &E_{1}^{c}\left(\mu_{0}, T_{0}\right) \nonumber\\
 &=&\frac{4\pi}{h^{3}}\int d^{3}r\int_{0}^{q_{c}}dp\,p^{2}\,\left(\frac{p^{2}}{2m}+U\left({\bf r}\right)\right)f_{0}^{\left(0\right)}\left(\epsilon;\mu_{0},T_{0}\right) \nonumber\\
 &=& \frac{\mathrm{e}^{-\eta_{0}}}{\lambda_{0}^{2}}\left\{ \frac{3\pi\hbar^{2}}{m}\frac{\mathrm{erf}\left[\sqrt{\eta_{0}}\right]}{\lambda_{0}^{3}\mathrm{e}^{-\eta_{0}}}-\frac{3q_{c}\hbar}{m\lambda_{0}^{2}}-\frac{q_{c}^{3}}{2\pi m\hbar}\right\} \int d^{3}r\, \mathrm{e}^{\alpha_{0}\left(\mathbf{r}\right)}\nonumber\\
 &+& \frac{\mathrm{e}^{-\eta_{0}}}{\lambda_{0}^{2}}\left\{ \frac{\mathrm{erf}\left[\sqrt{\eta_{0}}\right]}{\lambda_{0}\mathrm{e}^{-\eta_{0}}}-\frac{q_{c}}{\pi\hbar}\right\} \int d^{3}r\,U\left({\bf r}\right)\mathrm{e}^{\alpha_{0}\left(\mathbf{r}\right)},
\end{eqnarray}
and, for the quantum distributions,
\begin{eqnarray}\label{eq:027}
    & & E_{1}^{c}\left(\mu_{0},T_{0}\right) \nonumber\\
    &=&\frac{4\pi}{h^{3}}\int d^{3}r\int_{0}^{q_{c}}dp\,p^{2}\,\left(\frac{p^{2}}{2m}+U\left({\bf r}\right)\right) f_{0}^{\left(\pm\right)}\left(\epsilon;\mu_{0},T_{0}\right) \nonumber \\
    &=& \frac{3\pi\hbar^{2}}{m}\frac{1}{\lambda_{0}^{5}} \sum_{j=1}^{\infty}\left(\pm 1\right)^{j+1}\int d^{3}r\,\frac{{\rm e}^{j\alpha_{0}\left({\bf r}\right)}}{j^{5/2}}\mathrm{erf}\left[\sqrt{j\eta_{0}}\right] \nonumber\\
    &-&\frac{3q_{c}\hbar}{m}\frac{1}{\lambda_{0}^{4}} \sum_{j=1}^{\infty}\left(\pm 1\right)^{j+1}\int d^{3}r\,\frac{{\rm e}^{j\left(\alpha_{0}\left({\bf r}\right)-\eta_{0}\right)}}{j^{2}} \nonumber\\
    &-&\frac{q_{c}^{3}}{2\pi m\hbar}\frac{1}{\lambda_{0}^{2}}\sum_{j=1}^{\infty}\left(\pm 1\right)^{j+1}\int d^{3}r\,\frac{{\rm e}^{j\left(\alpha_{0}\left({\bf r}\right)-\eta_{0}\right)}}{j} \nonumber\\
    &+&\frac{1}{\lambda_{0}^{3}}\sum_{j=1}^{\infty}\left(\pm 1\right)^{j+1}\int d^{3}r\,U\left({\bf r}\right)\frac{{\rm e}^{j\alpha_{0}\left({\bf r}\right)}}{j^{3/2}}\mathrm{erf}\left[\sqrt{j\eta_{0}}\right]\nonumber\\
    &-&\frac{q_{c}}{\pi\hbar}\frac{1}{\lambda_{0}^{2}}\sum_{j=1}^{\infty}\left(\pm 1\right)^{j+1}\int d^{3}r\,U\left({\bf r}\right)\frac{{\rm e}^{j\left(\alpha_{0}\left(\mathbf{r}\right)-\eta_{0}\right)}}{j}.
\end{eqnarray}
Here, $\lambda_0=\left(h^{2}/2\pi mk_{B}T_0\right)^{1/2}$ is the thermal de Broglie wavelength at temperature $T_0$. After the system redistributes its energy $E_1$, it rethermalizes at a new temperature $T_1$. The system then acquires a new chemical potential, $\mu_1$, which determines the number of particles $N_1$. The gas thus attains a new equilibrium state, since the energy removed is significantly smaller than the total energy. At this stage, the distribution becomes $f_{1}^{\left(a\right)}\left(\epsilon; \mu_1, T_1\right)$, and the relevant expressions can be rewritten as follows:
\begin{equation}\label{eq:028}
N_{1}^{c}\left(\mu_0,T_0\right)=N_{1}\left(\mu_1,T_1\right),\quad E_{1}^{c}\left(\mu_0,T_0\right)=  E_{1}\left(\mu_1,T_1\right).
\end{equation}
Both the energy and the number of particles are functions of $T_1$ and $\mu_1$, which implies that these expressions can be inverted. Although this inversion is analytically challenging, it is more tractable numerically. Consequently, the following relations are obtained:
\begin{equation}\label{eq:029}
    T_{1} =  T_{1}\left(N_1,E_1\right), \quad \mu_{1} = \mu_{1}\left(N_1,E_1\right).
\end{equation}
These constitute a $2\times2$ nonlinear system of equations that can be solved for $T_1$ and $\mu_1$. All equations explicitly depend on the extensive variable $\mathcal{V}$; however, since the potential is considered fixed, this variable is eliminated when the equations are rewritten in terms of the initial quantities. This aspect will be addressed in the following sections.

\subsection{Applied evaporative cooling protocol \label{subsec:app-protocol}}

A critical aspect of implementing evaporative cooling is the conversion of Eqs. (\ref{eq:023})--(\ref{eq:027}) into a discretized and recursive form. This transformation establishes a systematic connection between the initial thermodynamic variables, namely the energy ($E_0$), particle number ($N_0$), chemical potential ($\mu_0$), and temperature ($T_0$), and their corresponding final values ($E_1$, $N_1$, $\mu_1$, $T_1$) after a cooling step, while ensuring that the appropriate functional form of the potential is selected\footnote{As defined in Sec. \ref{subsec:thermoconfining}, which satisfies the required properties.}. 

The procedure begins by specifying the initial thermodynamic state of the system, given by $(N_0, E_0, T_0, \mu_0)$, together with an initial cut-off momentum $q_c$. This cut-off defines an initial energy threshold $\eta_0$, above which particles are removed. To simulate evaporative cooling, particles with energies exceeding the reduced cut-off value $\eta_1 = \eta_0 - \Delta\eta$ are removed, resulting in a controlled linear decrease in temperature. After particle removal, the system rethermalizes and reaches a new equilibrium state characterized by updated thermodynamic quantities $(N_1, E_1, T_1, \mu_1)$. 

This cycle of lowering the cut-off energy, removing particles, and rethermalizing is repeated iteratively. The process continues until a target critical temperature $T_c$ is reached, as in the case of bosonic gases undergoing Bose–Einstein condensation, or until the particle population is depleted. The resulting discretized and recursive scheme provides a robust framework for modeling evaporative cooling, capturing both the progressive removal of high-energy particles and the corresponding evolution of thermodynamic properties at each stage.

The recurrence relations obtained generalize the computation of thermodynamic quantities for a gas that has lost its highest-energy particles in the $i$-th step, based on the values from the preceding $(i-1)$-th step, with the recursion initialized at step zero. Once the explicit forms of Eqs. (\ref{eq:023})--(\ref{eq:027}) are established, the protocol applies to any distribution function, classical or quantum. Here, it is implemented for the trapping potentials listed in Table \ref{tab:traps}, considering systems confined by a 3D box ($s=3/2$), a 3D harmonic oscillator ($s=3$), and a quadrupole trap ($s=9/2$). This analysis yields the following general recurrence relations for a classical gas:
\begin{equation}\label{eq:030}
    \frac{N_{i}}{N_{i-1}} = \mathrm{erf}\!\left(\sqrt{\eta_{i-1}}\right)
    - \frac{2}{\sqrt{\pi}}\,\eta_{i-1}^{1/2}\mathrm{e}^{-\eta_{i-1}},
\end{equation}
\begin{equation}\label{eq:031}
    \frac{E_{i}}{E_{i-1}} = C_{\mathrm{trap}}\,\mathrm{erf}\!\left(\sqrt{\eta_{i-1}}\right)
    - \frac{2}{\sqrt{\pi}}\,\eta_{i-1}^{1/2}\mathrm{e}^{-\eta_{i-1}}
    - \frac{2}{s\sqrt{\pi}}\,\eta_{i-1}^{3/2}\mathrm{e}^{-\eta_{i-1}}.
\end{equation}
Here, $C_{\mathrm{trap}} = 1 + \tfrac{3}{2s}$ is a geometrical factor associated with the trapping potential (see Appendix \ref{app:geometricalfactor}). These results agree with those reported for classical gases by Henn et al. \cite{Henn-amjphys75} and Davis et al. \cite{Davis-appphysb60}.

The recurrence relations for quantum gases are more complex but can be expressed in compact form:
\begin{equation}\label{eq:032}
    \frac{N_{i}}{N_{i-1}}=\frac{\tilde{g}^{\left( \pm \right)}_{s}\left(\alpha_{i-1},\eta_{i-1}\right)}{g^{\left( \pm \right)}_{s}\left(\alpha_{i-1}\right)}-\frac{2}{\sqrt{\pi}}\eta_{i-1}^{1/2}\frac{\bar{g}^{\left( \pm \right)}_{s-1/2}\left(\alpha_{i-1},\eta_{i-1}\right)}{g^{\left( \pm \right)}_{s}\left(\alpha_{i-1}\right)},
\end{equation}
\begin{eqnarray}\label{eq:033}
   \frac{E_{i}}{E_{i-1}} & = & \frac{\tilde{g}^{\left( \pm \right)}_{s+1}\left(\alpha_{i-1},\eta_{i-1}\right)}{g^{\left( \pm \right)}_{s+1}\left(\alpha_{i-1}\right)}-\frac{2}{\sqrt{\pi}}\eta_{i-1}^{1/2}\frac{\bar{g}^{\left( \pm \right)}_{s+1/2}\left(\alpha_{i-1},\eta_{i-1}\right)}{g^{\left( \pm \right)}_{s+1}\left(\alpha_{i-1}\right)} \nonumber\\
   & - & \frac{2}{s\sqrt{\pi}}\eta_{i-1}^{3/2}\frac{\bar{g}^{\left( \pm \right)}_{s-1/2}\left(\alpha_{i-1},\eta_{i-1}\right)}{g^{\left( \pm \right)}_{s+1}\left(\alpha_{i-1}\right)}.
\end{eqnarray}
For the results presented above, the following quantities are defined:
\begin{equation}\label{eq:034}
    \eta_{i}=\frac{\epsilon_c}{k_{B} T_{i}}=\frac{q_{c}^{2}}{2mk_{B}T_{i}}, \quad \alpha_{i}=\frac{\mu_{i}}{k_{B}T_{i}}.
\end{equation}
As previously noted, the recurrence relations do not depend on the extensive variable $\mathcal{V}$. However, to initialize the protocol, a specific value is selected for each potential $U\left({\bf r}\right)$ of interest. Additionally, $\bar{g}_{s}^{\left( \pm \right)}\left(\alpha,\sigma\right)$ and $\tilde{g}_{s}^{\left( \pm \right)}\left(\alpha,\sigma\right)$ are defined as modified Bose-Einstein and Fermi-Dirac functions\footnote{Similar to reported by Reyes-Ayala in ref. \cite{Reyes-Ayala2017}} (see Appendix \ref{app:gfunctions}). The bar function is given by:
\begin{equation}\label{eq:035}
\bar{g}_{s}^{\left( \pm \right)}\left(\alpha,\sigma\right)=\sum_{j=1}^{\infty}\left(\pm 1\right)^{j+1}\frac{{\rm e}^{j\left(\alpha
-\sigma\right)}}{j^{s}}.
\end{equation}
An important limiting case is $\lim_{\sigma\rightarrow\infty}\bar{g}_{s}^{\left( \pm \right)}\left(\alpha,\sigma\right)=0$. Similarly, for the tilde function:
\begin{equation}\label{eq:036}
\tilde{g}_{s}^{\left( \pm \right)}\left(\alpha,\sigma\right)=\sum_{j=1}^{\infty}\left(\pm 1\right)^{j+1}\frac{\mathrm{e}^{j \alpha }}{j^{s}}\mathrm{erf}\left[\sqrt{j\sigma}\right].
\end{equation}
In this case, $\lim_{\sigma\rightarrow\infty}\tilde{g}_{s}^{\left( \pm \right)}\left(\alpha,\sigma\right)=g_{s}^{\left( \pm \right)}\left(\alpha\right)$, which corresponds to the standard functions. This limiting behavior occurs because the error function rapidly approaches unity for large arguments.

\section{Numerical results and discussion}
\label{sec:numericalresults}

We apply the recursive protocol of Sec.~\ref{sec:evaporativecooling}, evaluated within the semiclassical framework of Sec.~\ref{sec:thermodynamics}, to the five geometries of Table~\ref{tab:traps} and to MB, BE, and FD statistics. All runs start from $N_0 = 1\times10^{7}$ atoms at $T_0 = 50~\mu$K with an initial cut-off $\eta_0 = 0.5$~mK, lowered in steps $\Delta\eta = 0.1~\mu$K. The confining geometry enters exclusively through the parameter $s$ of Table~\ref{tab:traps}, which simultaneously fixes the polylogarithm order, the density-of-states exponent, the power of $T$ in the particle-number prefactor, and the number of degrees of freedom $2s$. This single quantity organizes the entire discussion: it sets the ordering of the curves in every observable considered below, from the particle-number curves to the response functions.

\subsection{Evaporation routine}
\label{subsec:evaporative_routine}

\begin{figure*}[htpb!]
\centering
\includegraphics[width=0.75\linewidth]{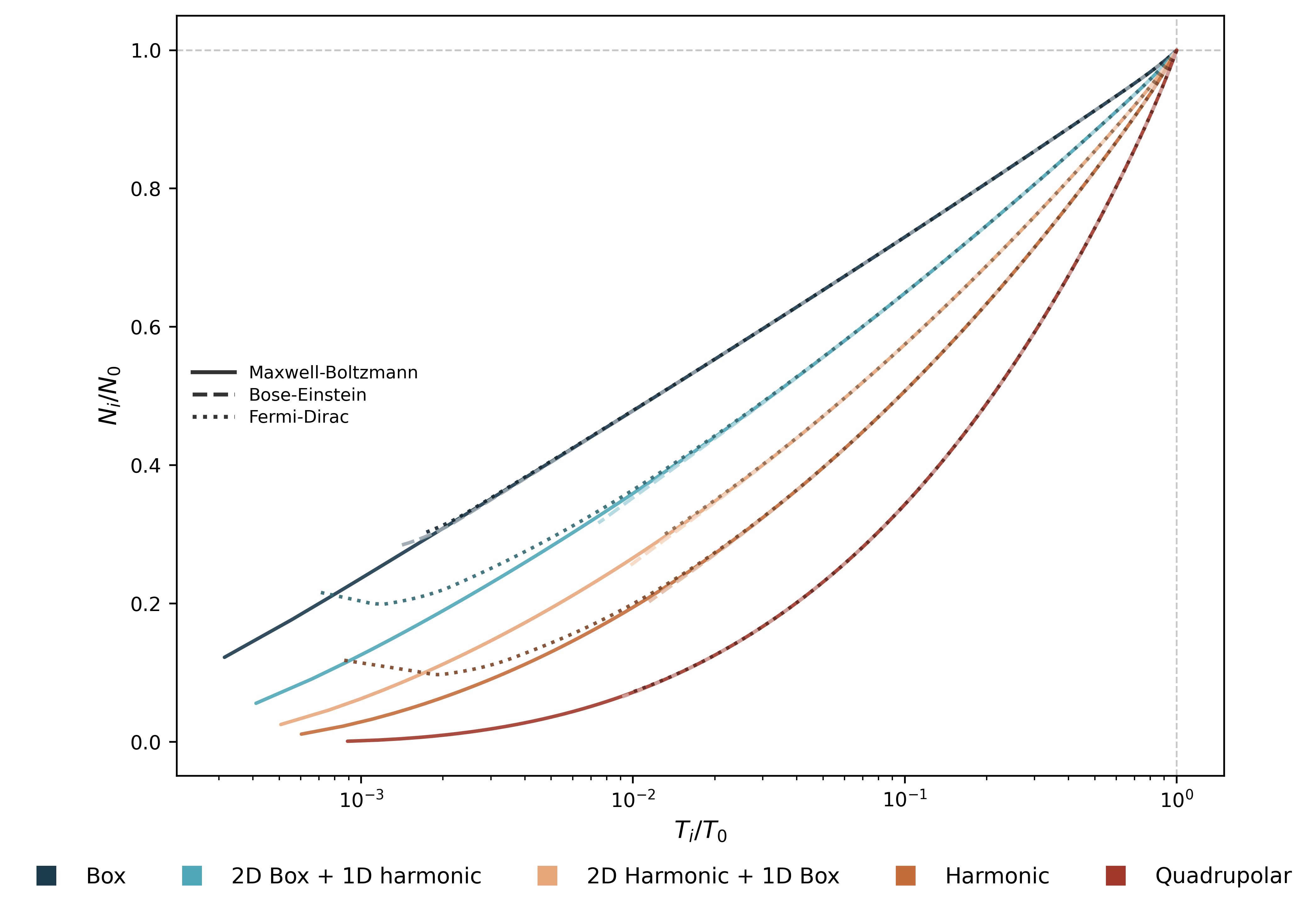}
\caption{\label{fig:evap_results}Evaporation curves in the normalized particle-number–temperature plane, $N_i/N_0$ versus $T_i/T_0$, are shown for the five geometries of Table \ref{tab:traps} and the three statistics (Maxwell–Boltzmann, solid; Bose–Einstein, dashed; Fermi–Dirac, dotted). All runs start from $N_0 = 1\times10^{7}$ atoms at $T_0 = 50~\mu$K with an initial cut-off $\eta_0 = 0.5$~mK lowered in steps $\Delta\eta = 0.1~\mu$K. At high temperatures, the curves coincide in the classical limit. Near degeneracy at low temperatures, the statistics separate: BE traces halt at condensation, while FD trajectories bend upward due to Pauli exclusion.}
\end{figure*}

Figure~\ref{fig:evap_results} shows the normalized particle number $N_i/N_0$ as a function of the reduced temperature $T_i/T_0$. Three regimes are apparent: a common classical branch at high temperature, a geometry-controlled ordering, and a low-temperature regime in which quantum statistics split the protocol.
 
At high temperature, every curve follows the same classical curve shape. Here $\alpha = \mu/k_B T$ is large and negative, the polylogarithmic corrections in equations~\eqref{eq:004}--\eqref{eq:006} are exponentially small, and the recurrences of equations~\eqref{eq:030}--\eqref{eq:032} reduce to their MB form. The coincidence of the MB, BE, and FD curves for every geometry confirms that the semiclassical distributions reproduce the classical limit irrespective of the confining potential.
 
The separation among geometries is already present in the classical branch, although its origin differs from the quantum splitting that appears later. The classical particle-number recurrence does not carry explicit $s$; the accompanying temperature reduction does, through the virial factor $C_{\mathrm{trap}} = 1 + 3/(2s)$ that weights the energy recurrence (Appendix~B). The classical traces therefore order monotonically with $s$, from the 3D box ($s = 3/2$, $C_{\mathrm{trap}} = 2$), which retains the largest atom fraction, to the quadrupole trap ($s = 9/2$, $C_{\mathrm{trap}} = 4/3$), which depletes fastest. The larger $s$ thus corresponds to a higher particle loss per unit of temperature reduction, consistent with the energy scaling $E = s N k_B T$ of Eq.~\eqref{eq:020}. Quantum statistics preserve this ordering and amplify it near degeneracy, where the density of states enters directly.
 
Quantum-statistical effects dominate the low-temperature end. For BE statistics, each trajectory terminates when $\mu \to 0$: the Bose function $g_s^{(-)}(\alpha)$ saturates at $\zeta(s)$, the temperature can no longer be lowered, and $N_i/N_0$ plateaus at the onset of condensation. The critical temperatures obtained for the box, harmonic, and quadrupole traps reproduce the ideal-gas values, validating the numerical implementation. The quadrupole trap condenses in the idealized model, but its experimental realization is precluded by Majorana losses at the field zero, as noted in Sec.~\ref{sec:thermodynamics}. For FD statistics, the trajectories bend upward: because the low-lying states are already occupied, removing high-energy atoms forces the remainder into higher single-particle levels, raising the mean energy per particle \cite{Mendoza-Lopez2016}. This Pauli-driven effective heating is a direct signature of Fermi degeneracy, and the BE--FD separation widens with $s$ as the density of states amplifies the statistical contrast.

\subsection{Thermodynamic potentials}
\label{subsec:thermodynamic_potentials}

\begin{figure*}[htpb!]
\centering
\includegraphics[width=1.00\linewidth]{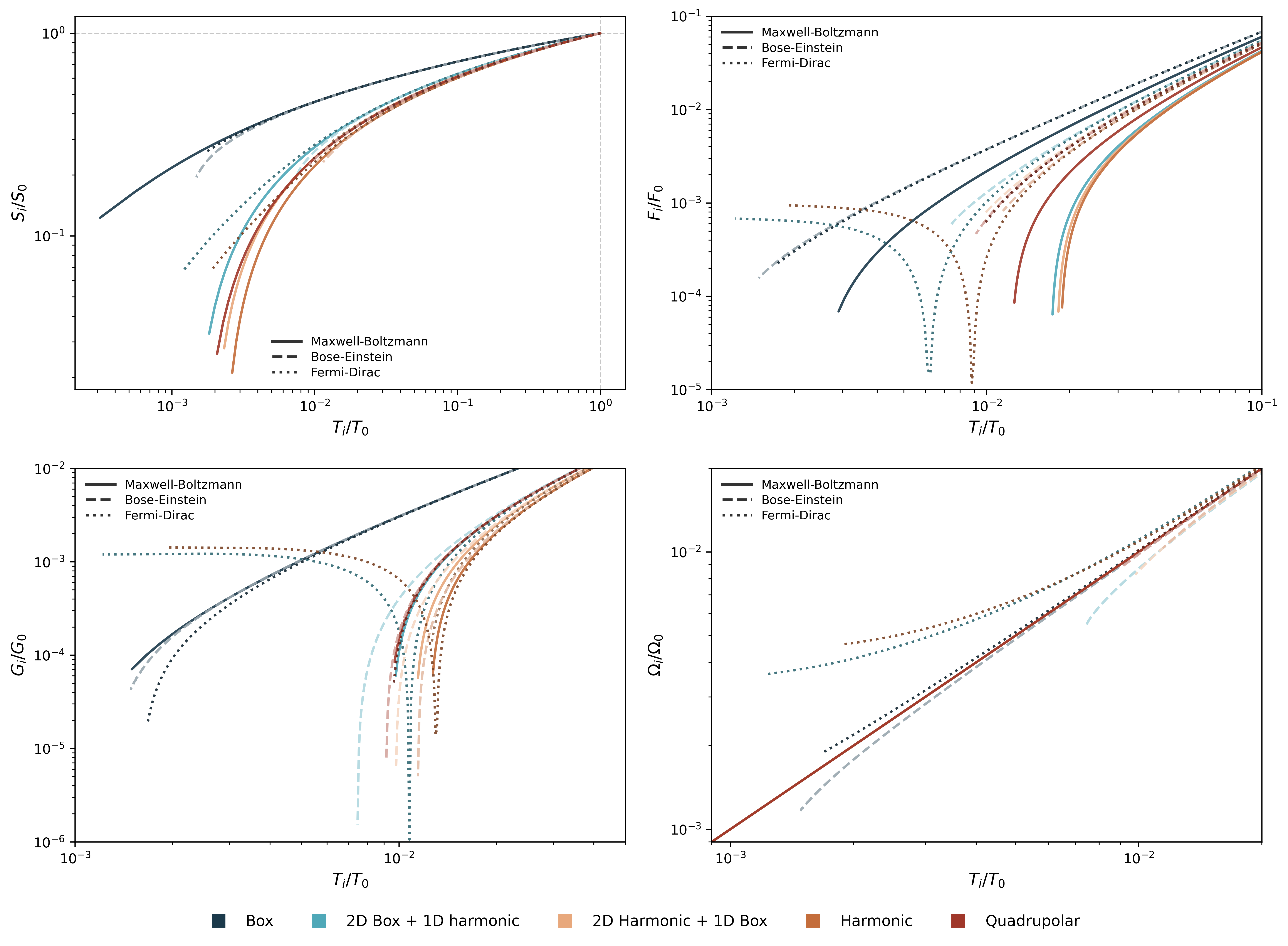}
\caption{\label{fig:free_energies}Specific thermodynamic potentials $s_i/s_0$, $f_i/f_0$, $g_i/g_0$, and $w_i/w_0$ versus the reduced temperature $T_i/T_0$ for the five geometries and three statistics. The classical and quantum branches coincide at high temperatures, with quantum deviations emerging near the degeneracy point. Cusps in $f$ and $g$ arise when the chemical potential crosses zero. In the entropy panel, FD lies below BE due to Pauli suppression, while BE curves terminate at condensation.}
\end{figure*}

Figure~\ref{fig:free_energies} shows the specific entropy, Helmholtz free energy, Gibbs free energy, and grand potential, $s_i/s_0$, $f_i/f_0$, $g_i/g_0$, and $w_i/w_0$, versus $T_i/T_0$. All branches again coincide on the classical limit at high temperature and separate near degeneracy; we focus on the feature that distinguishes each potential.

The entropy panel isolates the contrast between the two quantum statistics. The FD curves lie below their BE counterparts throughout the degenerate regime, because Pauli exclusion suppresses low-energy excitations, whereas bosonic bunching preserves the entropy down to the condensation threshold, where the BE curves terminate. The geometric ordering follows the degrees of freedom $2s$.
 
The Helmholtz and Gibbs panels develop sharp downward cusps. As the chemical potential crosses zero, $f$ and $g$ change sign, and since their magnitudes are plotted, each curve is driven to zero at the crossing. The cusp thus marks the onset of degeneracy, and its temperature increases with $s$: steeper traps reach $\mu = 0$ first, in agreement with the particle-number trace of Figure~\ref{fig:evap_results}.
 
The grand potential reproduces the same ordering. Because $w$ generates the global variables and susceptibilities of Section~\ref{sec:thermodynamics}, its behavior anticipates the response functions discussed below: the BE curves terminate at condensation, while the FD curves flatten as the Pauli-blocked gas reorganizes its occupation of the low-lying states.

\subsection{Susceptibilities}
\label{subsec:susceptibilities}

\begin{figure*}[htpb!]
\centering
\includegraphics[width=1.00\linewidth]{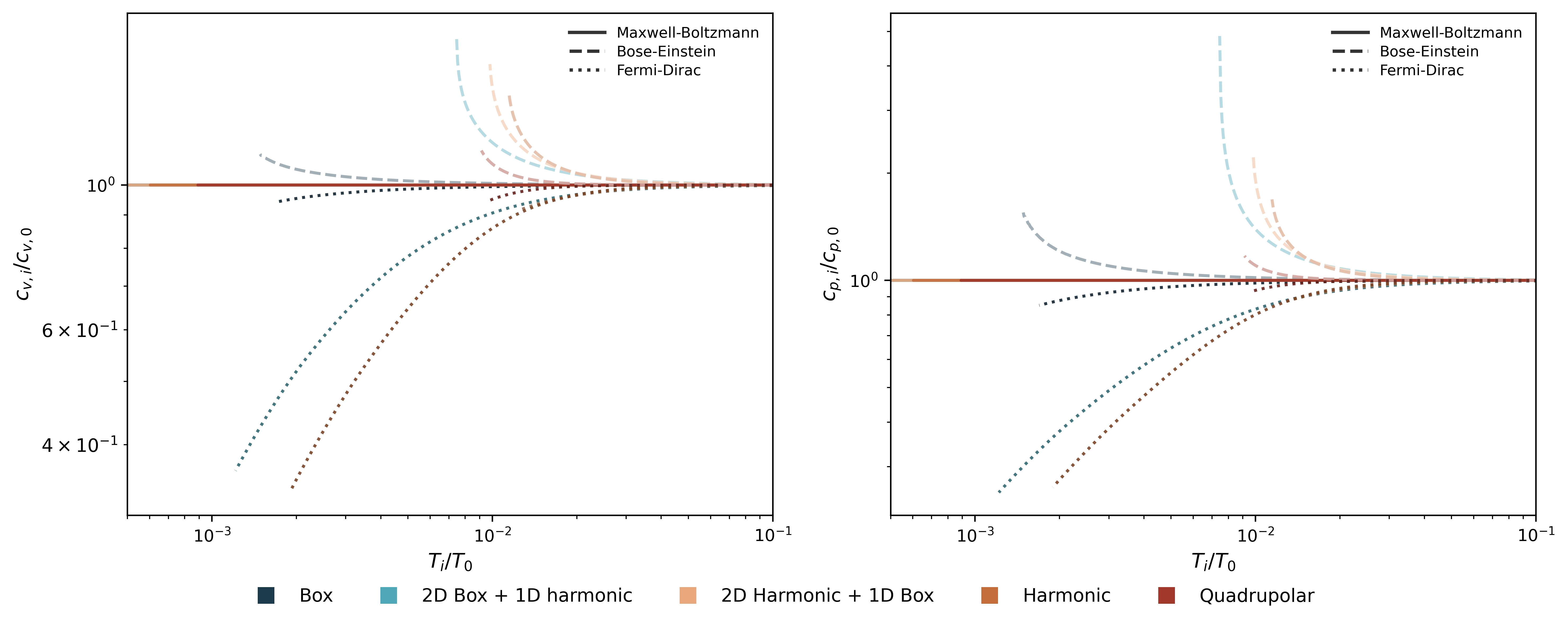}
\caption{\label{fig:heat_capacities}Heat capacities $c_{\mathcal{V},i}/c_{\mathcal{V},0}$ and $c_{\mathcal{P},i}/c_{\mathcal{P},0}$ versus $T_i/T_0$ for the five geometries and three statistics. MB curves remain at the classical plateau. Near degeneracy, BE capacities rise and peak at condensation, while FD capacities drop below the classical value due to Pauli suppression. The departure temperature is set by the trap density of states, with larger $s$ separating earlier.}
\end{figure*}

\begin{figure*}[htpb!]
\centering
\includegraphics[width=1.00\linewidth]{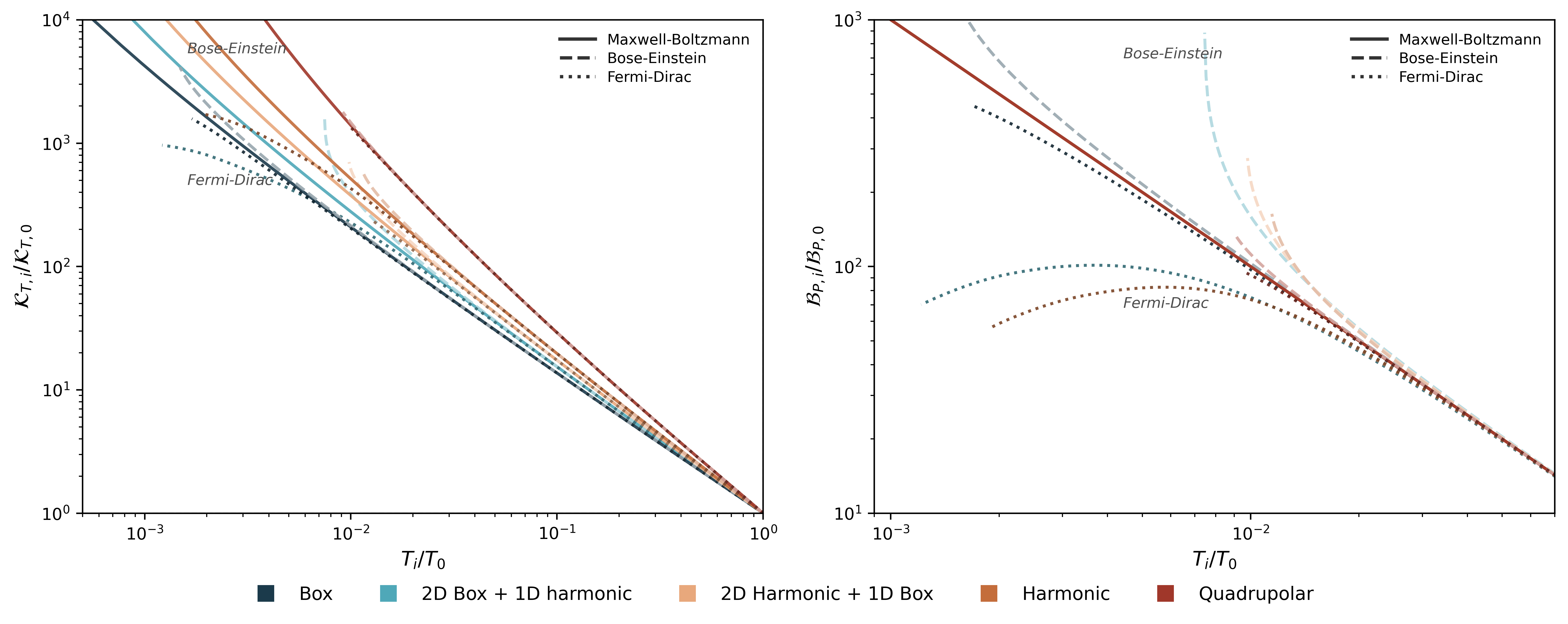}
\caption{\label{fig:compresisbilities} The left panel shows the normalized isothermal compressibility $\kappa_{T, i}/\kappa_{T, 0}$ and the right panel the response $\mathcal{B}_{P, i}/\mathcal{B}_{P,0}$ versus $T_i/T_0$ for Maxwell–Boltzmann, Bose–Einstein, and Fermi–Dirac gases. All curves coincide with the classical power law at high temperatures. As $T$ decreases, quantum degeneracy produces strong departures from classical behavior, with thresholds and scaling set by the trap density of states.}
\end{figure*}

Figs.~\ref{fig:heat_capacities} and~\ref{fig:compresisbilities} show the response functions versus $T_i/T_0$, each normalized so that the classical MB plateau equals unity. Two mechanisms organize their low-temperature behavior. Bosonic softening toward the ground state near the critical temperature enhances the response and, in the heat capacities, produces a peak at the condensation temperature. Fermionic stiffening has the opposite effect: as a rigid Fermi sea forms, Pauli exclusion pushes particles to higher momenta and generates a Fermi pressure that suppresses the response below its classical value. The temperature at which each curve leaves the classical plateau is set by the density of states through $s$, so that steeper traps depart earliest and box-like geometries remain classical over the widest range. On logarithmic axes, the classical branches are linear, confirming the expected power-law scaling of the response functions.

The heat capacities $c_{\mathcal{V}}$ and $c_{\mathcal{P}}$ (Fig.~\ref{fig:heat_capacities}) follow this pattern directly. The BE curves rise sharply and peak at condensation, whereas the FD curves fall below unity. These trends mirror, respectively, the entropy preserved by bosonic bunching and suppressed by Pauli exclusion in Fig.~\ref{fig:free_energies}, and the corresponding BE and FD particle-number trajectories of Fig.~\ref{fig:evap_results}.

The isothermal compressibility $\kappa_{T}$ and the global thermal expansion coefficient $\mathcal{B}_{P}$ (Fig.~\ref{fig:compresisbilities}) probe the mechanical response. For fermions, the Fermi pressure saturates $\kappa_{T}$ at a finite value, the mechanical signature of the incompressible Fermi sea. For bosons, the compressibility diverges in box-like geometries as the gas softens toward condensation; harmonic confinement reshapes the density of states and moderates this divergence. The thermal expansion coefficient deviates from the classical power law over the same temperature range, with enhancement for bosons and suppression for fermions, in a manner that tracks the compressibility. The response functions thus provide the sharpest thermodynamic markers of the normal-to-degenerate crossover: its location is fixed by the geometry, while the sign and shape of the departure are fixed by the statistics.

\section{Conclusions}

We have developed a semiclassical framework that unifies the evaporative cooling of classical and quantum gases across a broad family of confining potentials. By introducing generalized thermodynamic variables and deriving recurrence relations for the particle number and internal energy, we track the full thermodynamic state of the gas at every step of the cooling protocol. The confining geometry enters through the single parameter $s$, which fixes the polylogarithm order, the density-of-states exponent, and the number of degrees of freedom $2s$, and which classically also controls the energy budget through the virial factor $C_{\mathrm{trap}} = 1 + 3/(2s)$.
 
The trajectories of Figs.~\ref{fig:evap_results}--\ref{fig:compresisbilities} show that trap geometry and quantum statistics jointly determine the efficiency and endpoint of evaporation. In the classical limit, the gas depletes smoothly and completely, with the geometry ordering set by $C_{\mathrm{trap}}$. For bosons, the particle-number trajectories, thermodynamic potentials, and susceptibilities all terminate or peak at condensation, where the Bose function saturates. For fermions, Pauli blocking produces a coherent set of degeneracy signatures, the upward bending of the $N$--$T$ curves, the suppressed entropy and heat capacities, and the saturating compressibility, all traceable to the effective heating of the truncated distribution. Because $s$ sets the density of states, these signatures appear earliest and are most pronounced in the quadrupole trap ($s = 9/2$), whose nonstandard phase-space scaling makes it the most sensitive probe of degeneracy among the geometries considered; its condensation, however, is an idealization not accessible experimentally owing to Majorana losses.
 
Among the observables, the response functions provide the sharpest markers of the normal-to-degenerate crossover: the crossover temperature is fixed by the geometry, whereas the sign and shape of the departure from classical behavior are fixed by the statistics. Although interactions are neglected, the polylogarithmic structure of the BE and FD distributions reproduces the equilibrium behavior of the trapped gas and captures how the confining potential, through $s$, controls both the rate of temperature loss and the evolution of the particle number during cooling. The framework thus offers a compact and experimentally grounded description of evaporative cooling across relevant trapping geometries, and provides a natural starting point for extensions that include interactions, non-equilibrium dynamics, or time-dependent protocols.

\bibliography{bib_semiclassical_evap_cooling}

\end{document}